\documentclass[11pt, oneside]{amsart}

\usepackage{amsmath}
\usepackage{amssymb}

\usepackage{color}
\usepackage{dcolumn}
\usepackage{float}
\usepackage{graphicx}
\usepackage[latin9]{inputenc}
\usepackage{multirow}
\usepackage{rotating}
\usepackage{subfigure}
\usepackage{psfrag}
\usepackage{tabularx}
\usepackage[hyphens]{url}
\usepackage{wrapfig}
\usepackage{framed,color}
\usepackage{fancybox}
\usepackage{verbatim}

\usepackage[bookmarks, bookmarksopen, bookmarksnumbered]{hyperref}
\usepackage[all]{hypcap}
\definecolor{orange}{rgb}{1.0,0.3,0.0}
\definecolor{violet}{rgb}{0.75,0,1}
\definecolor{darkgreen}{rgb}{0,0.6,0}
\definecolor{cyan}{rgb}{0.2,0.7,0.7}
\definecolor{blueish}{rgb}{0.2,0.2,0.8}

\definecolor{shadecolor}{rgb}{0.9,0.9,0.9}

\DeclareRobustCommand{\csdms}{\textsc{csdms}}
\DeclareRobustCommand{\doi}{\textsc{doi}}
\DeclareRobustCommand{\bmi}{\textsc{bmi}}
\DeclareRobustCommand{\wmt}{\textsc{wmt}}

\begin{document}

\title[]{Building Sustainable Software - The CSDMS Approach}

\author{
  Eric W. H. Hutton$^{\dag}$,
  Mark D. Piper$^{\dag}$,
  Scott D. Peckham$^{\dag}$,
  Irina Overeem$^{\dag}$,
  Albert J. Kettner$^{\dag}$,
  James P. M. Syvitski$^{\dag}$,
}

\thanks{{}$^{\dag}$Community Surface Dynamics Modeling System, University of
Colorado}

\begin{abstract}

\csdms{}, The Community Surface Dynamics Modeling System, is an NSF funded
project whose focus is to aid a diverse community of earth and ocean system
model \emph{users} and \emph{developers} to use and create robust software
quickly.  To this end, \csdms{} develops, integrates, archives and disseminates
earth-system models and tools to an international (67 country) community
with the goal of building the set of tools necessary to model the
earth system. Modelers use \csdms{} for access to hundreds of open
source surface-dynamics models and tools, as well as model metadata. Such a
model repository increases model transparency and helps eliminate
duplication by presenting the current state of modeling efforts.
To increase software sustainability, composability and interoperability,
\csdms{} promotes standards that define common modeling interfaces, semantic
mediation between models, and model metadata. Through online resources and
workshops, \csdms{} promotes software engineering best practices, which are
unfamiliar to many developers within our modeling community. For example,
version control, unit testing, continuous integration, test-driven
development, and well-written \emph{clean code} are all topics of the
educational mission of \csdms{}.

\end{abstract}

\maketitle

\section{The Community Surface Dynamics Modeling System}

The mission of the Community Surface Dynamics Modeling System (\csdms{})
~\cite{peckham2012component} is to help a diverse modeling community toward
common goals and standards. This effort involves creating:
\begin{itemize}
  \item a repository of source code and metadata for open-source models and
    tools (Section~\ref{sec:organizing})
  \item reusable plug-and-play model components and a framework
        within which they can be coupled to create new models
        (Section~\ref{sec:reusable})
  \item an \emph{efficient} and \emph{open} modeling community through
        standards (Section ~\ref{sec:standards}) and education
        (Section~\ref{sec:education})
\end{itemize}

In building a modeling framework, \csdms{} has leveraged
several existing, well-established and open-source software tools. For example,
\csdms{} uses tools from the Common Component Architecture (CCA)
~\cite{armstrong1999toward} toolchain: \emph{Babel}, and \emph{Bocca}. Babel
provides interoperability between components
written in different languages; it currently supports C, C++, Fortran, Java,
and Python. Bocca helps with creating CCA-compliant components and managing CCA
component projects.

\csdms{} has developed innovative model/component interfaces that promote
model reuse and interoperability,
including the Basic Model Interface (\bmi{} - Section~\ref{sec:bmi}), which
uses
the \csdms{} Standard Names (Section~\ref{sec:standardnames}) and a framework
within which models can be coupled and enhanced with a set of \bmi{}-compatible
service components.
The \csdms{} Web Modeling Tool (\wmt{} - Section~\ref{sec:wmt}) is a web-based application that provides a
graphical interface to this framework and allows users to compose new models
by connecting and configuring components in a simple, browser-based graphical
interface.

\section{Organizing and Documenting Open-Source Models}
\label{sec:organizing}

\subsection{The \csdms{} Model Repository}
\label{sec:repository}

Open-source software reduces redundancy, provides transparency through
external review and makes research replicable, which is fundamental to
scientific practices~\cite{ince2012case}.
Software accessibility is a key value of the \csdms{} mission. \csdms{}
ensures code developed by individual researchers or small research teams
remains accessible beyond the lifetime of a specific research project. Code
developers are required to distribute their codes under open-source licenses. 

\csdms{} has built an online model repository that not only contains model
source code but also documents the submitted code as model metadata, runs
basic smoke-tests of the models, (optionally) maintains the source code in a
\csdms{}-hosted version control system, and provides easy downloads of the
original source. The repository now contains over 200 models and tools (over
4M lines of code) and has received over 13k model downloads.

\subsection{Digital Object Identifiers for Numerical Models}
\label{sec:doi}

\csdms{} has implemented a
Digital Object Identifier (\doi{}) system for software within its model
repository to guarantee recognition to software contributors.
\csdms{} was among the first software venues to assign \doi{}s to
open source software. The advantages of adopting a \doi{} system for open
source software are: 
\begin{itemize}
\item Guarantee credit to the software developer
\item Easily reuse and replicate research as software is directly locatable 
\item Increase software visibility: \doi{} content is 5 times more likely to
      deliver active links than content without.
\item Provide funding agencies with the ability to track usage, to measure
      impact.
\end{itemize}

Only stable versions of open source
software that are physically hosted through the \csdms{} repository will be
assigned a \doi{}. For model software citations, \csdms{} recommends the
following the DataCite guidelines~\cite{brase2009datacite}.

\begin{shaded}
\leftskip 0.25in
\parindent -0.25in
\tt{
Developer, A., Developer, B. (Year of publication). \emph{Name of the model},
Model Version. Identifier.
}
\end{shaded}

\section{Reusable Components}
\label{sec:reusable}

\subsection{Sustainability and Interoperability Through Standards}
\label{sec:standards}

\subsubsection{The Basic Modeling Interface}
\label{sec:bmi}

The \emph{Basic Modeling Interface}\footnote{https://github.com/csdms/bmi}
(\bmi{}) is a component-coupling library
interface specification designed by \csdms~\cite{peckham2012component,
syvitski2014plug}.  In this context, a component is software that models a
particular environmental process and can be \emph{plugged into} another
component.

The \bmi{} specification was designed without any particular model-coupling
framework in mind.  Rather, it was designed to be framework agnostic. We
recognize model-coupling frameworks and tools may come and go but the
functionality a framework requires from its components will be significantly
more long-lived. Thus, a component should outlive any framework that it might
operate within.

The \bmi{} identifies the entry points into software components that provide a
calling application with the necessary level of control to connect components.
\csdms{}, as well as other modeling frameworks such as
ESMF~\cite{hill2004architecture}, OpenMI~\cite{gregersen2007openmi}, and
OMS~\cite{david2002object}, identifies an interface that, at a minimum,
provides functionality to \emph{initialize}, \emph{update}, and
\emph{finalize} a component. For more complicated (and 2-way)
coupling, components must implement \bmi{} methods providing data access,
and model metadata. For data access, \bmi{} defines a set of \emph{getter}
and \emph{setter} methods that allow model components to present internal data
for both reading and writing.

\csdms{} has found that \bmi{} is an acceptable target for model developers.
The use of \bmi{} has also dramatically reduced the effort required by
\csdms{} staff to create and maintain components. So long as a model component
strictly exposes a \bmi{}, it can automatically be incorporated into the
\csdms{} model-coupling framework where it can then connect to compatible
components to form larger models and obtain additional functionality through
\bmi{} compatible tools (for example, NetCDF output, grid mapping).

\subsubsection{\csdms{} Standard Names}
\label{sec:standardnames}

The \csdms{} \emph{Standard Names}
\footnote{https://github.com/csdms/standard\_names}
are a common language for variable names exchanged between models. They play
an important role in the \bmi{} as they provide a mapping of a model's
internal variable names to a common language used by the \bmi{} getter and
setter functions.

Most models require input variables and produce output variables. For model
components to be reusable and interoperable (either within the \csdms{}
framework or any framework in general)
a set of components become a complete model when every component is able to
obtain the input variables it needs from another component. The \csdms{}
Standard Names were developed to provide a practical solution to this
semantic mediation problem ~\cite{peckham2012component, syvitski2014plug}.
The \csdms{} Standard Names provide a comprehensive set of naming rules and
patterns for creating unique labels for model variables that are not specific
to any particular modeling domain.

\subsection{Connecting and Running Models Through the Web}
\label{sec:wmt}

\begin{figure}
  \caption{The \csdms{} Web Modeling Tool}
  \begin{center}
    \includegraphics[scale=.25]{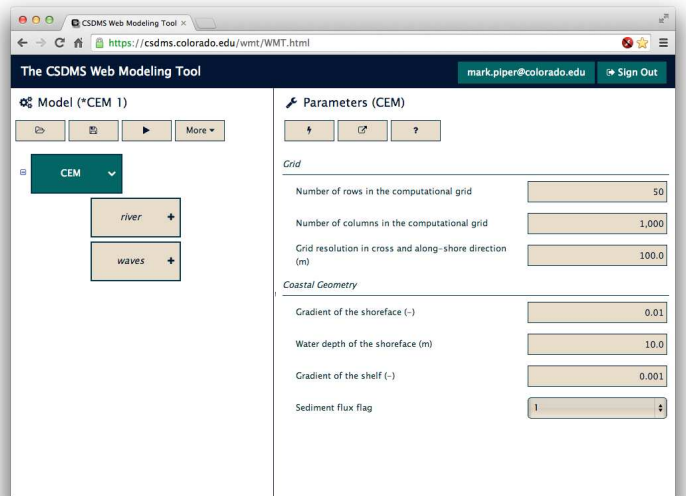}
  \end{center}
  \label{fig:wmt_screenshot}
\end{figure}

The \csdms{} Web Modeling Tool\footnote{https://csdms.colorado.edu/wmt}
(\wmt{}) is a web
application that provides an Ajax client-side (the \wmt{}
client\footnote{https://github.com/csdms/wmt-client}) graphical interface
(Figure~\ref{fig:wmt_screenshot}) and a RESTful
server-side database and API (the \wmt{} server\footnote{https://github.com/csdms/wmt}) that allows users to build and run coupled Earth system models on
a high-performance computing cluster from a web browser. \wmt{} was designed
with four objectives:
\begin{itemize}

\item \emph{Accessibility}. As a web-based application, users with internet
      access, have access to \wmt{}.

\item \emph{Integration}. Easily hyperlink from \wmt{} to resources on the 
      \csdms{} modeling portal (model documentation, labs, lectures,
      tutorials)

\item \emph{Portability}. \wmt{} has a native JavaScript interface.
      JavaScript is an open standard and almost every computer is equipped
      with a JavaScript-capable web browser.


\end{itemize}
With \wmt{}, users can: \emph{run} single component models, build (2-way)
\emph{coupled models}, view and \emph{edit} model parameters, and \emph{share}
saved models with others in the community.

\section{Education}
\label{sec:education}


It has been our experience that new graduate students in the Earth sciences
begin with little or no programming experience. However, their interest in
programming is high and they see the importance of programming with regard
to their thesis topics. More often than not, programming courses are not
offered as part of their coursework and so, out of necessity to solve a
particular problem, students end up teaching themselves how to program.

We believe providing these students with even a small base in programming and
guidelines on best practices will go a long way. We have hosted several
programming workshops, including one through Software Carpentry. The workshops
were such a success that we are now planning an annual Software Carpentry
bootcamp that is specifically targeted to incoming graduate students. The
workshops fill up nearly immediately, which reflects a strong interest and
need for basic software skills. We believe these workshops strike a good
balance between teaching students about computer science topics and
maintaining a pragmatic focus on problem solving.



\section{Summary: Building an active and efficient modeling community}

\csdms{} provides cyber-infrastructure to promote the quantitative modeling of
earth surface processes. The US National Science Foundation funds \csdms{}, but
the community is international and includes industry and federal agency
representatives in addition to members from academic institutions. The
initiative has been growing with more than 150 people per year. \csdms{} now
fosters a community of approximately 1200 scientists who work on prediction
of the movement of fluids, and the production, erosion, transport, and
deposition of sediment and nutrients in landscapes and seascapes. 

\section{Acknowledgements}

The \csdms{} Integration Facility operates under continuing grant 0621695
from the US National Science Foundation.

\bibliographystyle{plain}

\bibliography{wssspe-citations}{}

\end{document}